\documentclass[aps,prl,twocolumn,10pt]{revtex4}

\usepackage{verbatim}
\usepackage{amsmath}
\usepackage{amsfonts}
\usepackage{amssymb}
\usepackage{bbm}
\usepackage{latexsym}
\usepackage{hyperref}

\usepackage{subfig}
\usepackage{multirow}
\usepackage{mathrsfs}

\usepackage{amscd}

\begin{document}
%
%
\title{A Lorentz invariant doubled worldsheet theory}
\preprint{LMU-ASC 50/12, MPP-2012-118}
\author{Stefan \surname{Groot Nibbelink}$^{a}$, Peter \surname{Patalong}$^{a,b}$}
\affiliation{%
\vskip0.15cm
$^{a}$Arnold-Sommerfeld-Center for Theoretical Physics, Fakult\"at f\"ur Physik,  Ludwig-Maximilians-Universit\"at M\"unchen, Theresienstra\ss e 37, 80333 M\"unchen, Germany. \\ 
\\ \vskip-0.6cm 
$^{b}$Max-Planck-Institut f\"ur Physik, F\"ohringer Ring 6, 80805 M\"unchen, Germany. 
}%
\begin{abstract}
We propose a Lorentz invariant version of Tseytlin's doubled worldsheet theory that makes T--duality covariance of the string manifest. This theory can be derived as a gauge fixed version of Buscher's gauging procedure, in which the left--over gauge field component acts as a Lagrange multiplier. This description can naturally account for fractional linear $O(D,D)$ transformations of the metric and $b$--field. It is capable of describing non--geometric backgrounds; geometric and non--geometric fluxes are encoded in the doubled anti--symmetric tensor field strength. 

\end{abstract}
%
%
\maketitle

\section{Introduction}

T--duality is a fundamental symmetry between certain string backgrounds. Buscher's initial approach  \cite{b87a,b87b} implements it as a transformation of the target space metric and $b$--field, but does not realize it as a manifest symmetry on the worldsheet. On the contrary, Tseytlin's approach \cite{t90a,t90b} (based on \cite{Floreanini:1987as}, see also \cite{Duff:1989tf,Dall'Agata:2008qz}) realizes T--duality as a manifest $O(D,D)$ covariance of a worldsheet action for doubled coordinate fields. This description is not (manifestly) Lorentz invariant, which poses complications for interpreting the $\beta$-functions as target space equations of motion (e.o.m.'s) \cite{Sfetsos:2009vt,Avramis:2009xi}. 

The doubled formalism by Hull \cite{Hull:2004in,Hull:2006va} describes the same number of target space d.o.f.'s as the standard sigma--model because the constraint and the kinetic terms are assumed to be related. Ref.\ \cite{Berman:2007yf,Berman:2007xn} gave a Lorentz invariant description of the chirality constraint \cite{Pasti:1996vs}. This resulted in a non--linear theory unless one makes a gauge fixing, leading back to Tseytlin's form. In this letter, we propose a manifest Lorenz invariant doubled worldsheet theory which overcomes these issues.

Various approaches that double the coordinates have been utilized to investigate ``non--geometry''. We use our proposed formalism to analyze the prime example of a non-geometric background, namely the three-torus with a linear $b$--field and its T--duals. It turns out that in this case our formalism avoids the ambiguities that typically show up in any non--geometric setup. Furthermore, we obtain the chain of non--geometric fluxes as the field strength of the doubled anti--symmetric tensor field.

\section{T-duality on the worldsheet}

The starting point is the worldsheet path integral for the coordinate fields $X^\mu(\sigma)$, $\mu =0,1,\ldots D\mbox{-}1$,
\begin{equation}
Z = \int\!\mathcal{D}[X] \sqrt{\det E(X)}\, \exp\, 
i\! \int\! 
\partial_L X^T E(X)\, \partial_R X~, 
\label{Polyakov} 
\end{equation}
with $E(X)= g(X)+b(X)$, that map into a target space with metric $g(X) = [g_{\mu\nu}(X)]$ and anti--symmetric tensor $b(X) = [b_{\mu\nu}(X)]$.  We defined left-- and right--moving derivatives as $\partial_{L/R} = (\partial_0 \pm \partial_1)/\sqrt 2$ and dropped the worldsheet measure: $\int = \int\!d^2\sigma$.  
This theory, describing $D$ bosonic  d.o.f.'s, is both Lorentz and scale invariant. Requiring Weyl invariance (finiteness) at the loop level results in the target space e.o.m.'s \cite{Callan:1985ia,Hull:1985rc}. 

\subsection{Buscher's approach}

Buscher's approach \cite{b87a,b87b} assumes that the metric  $g$ and the Kalb-Ramond tensor $b$ are constant or at least possess some isometries. They are promoted to local symmetries, 
\begin{equation}
X(\sigma) \rightarrow X(\sigma) - \xi(\sigma)~,
\quad 
V_{a}(\sigma) \rightarrow V_{a}(\sigma) + \partial_{a}\xi(\sigma)~, 
\label{GaugeTrans} 
\end{equation}
by introducing a gauge connection $V_a$ with $a =0,1$ or $L,R$.  The covariant derivatives $D_{a} X = \partial_{a} X + V_a$ and field strength 
$F = \partial_R V_L - \partial_L V_R$ are inert under these gauge transformations. The number of worldsheet d.o.f.'s remains the same provided that a Lagrange multiplier $\tilde X = (\tilde X_\mu)$ enforces $V$ to be pure gauge $\smash{F\stackrel{!}{=}0}$.

The quantization requires a gauge fixing condition $\smash{G \stackrel{!}{=}0}$ implemented by a Lagrange multiplier $\tilde{B}$. Different choices for $G$ lead to equivalent descriptions of the same physics.  In the path integral 
\begin{align}
Z &= \int\!\mathcal{D}[X, \tilde X, V, \tilde{B}, \textsc{b}, \textsc{c}]
\sqrt{\det E}\, \exp i S~,
\label{PathInt} 
\\
S &= \int\! 
\Big\{ D_L X^T E\, D_R X 
+  \tilde{X}^T F 
+ \tilde{B}^T G 
+ \textsc{b}^T \delta_\textsc{c} G 
 \Big\}~
\nonumber
\end{align}
the Faddeev--Popov $(\textsc{b}, \textsc{c})$ ghost's action is determined by the variation of the gauge fixing condition $\delta_\xi G$ with the gauge parameter $\xi$ replaced by $\textsc{c}$.

Taking $G = X$ the ghosts can be integrated out trivially. The integrals over $\tilde{B}$ and $X$ implement the gauge fixing $\smash{X\stackrel{!}{=}0}$. After integrating the $V_a$'s the path integral only involves the dual coordinate fields $\tilde X$: 
\begin{equation}
Z = \int \mathcal{D}[\tilde{X}] 
\sqrt{\det \tilde{E}}\,  \exp\, i\! \int\! 
\partial_L \tilde{X}^T \tilde E\, \partial_R \tilde{X}~,  
\label{DualAction} 
\end{equation}
with $\tilde g + \tilde b = \tilde{E} = E^{-1} = (g + b)^{-1}$. 
This motivated the factor $\sqrt{\det E(X)}$ rather than $\sqrt{\det g(X)}$ in the path integral \eqref{Polyakov} because it not only ensures that it is invariant under target space diffeomorphisms but also transforms covariantly under T--duality. Hence, the Buscher's approach maps one worldsheet action to its T--dual version but does not make T--duality covariance manifest.

\subsection{Tseytlin's approach}

As briefly explained in an appendix of \cite{Rocek:1997hi} one can obtain a T--duality covariant description by enforcing the axial gauge: $G=V_1$. The integrals over $\tilde{B}$ and $V_1$ implement the gauge fixing $\smash{V_1\stackrel{!}{=}0}$. The integral over $V_0$ leads to a path integral of the action 
\begin{equation}
S = \int 
\Big\{
-\frac{1}{2}\, \partial_1 Y^T \mathcal{H}\, \partial_1 Y 
- \frac{1}{2}\, \partial_1 Y^T \eta\, \partial_0 Y 
+ \textsc{b}^T \partial_1 \textsc{c} 
\Big\}~, 
\label{Tseytlin}
\end{equation}
with $Y^T = \begin{pmatrix} X^T & \tilde{X}^T \end{pmatrix}$. Further integrating over the ghosts gives the functional determinant $\det \partial_1$ in the path integral as in \cite{t90b}. The matrices 
\begin{equation}
\eta = \begin{pmatrix} 
0 & \mathbbm{1} \\ \mathbbm{1} & 0 
\end{pmatrix}~,~
\mathcal{H} = \begin{pmatrix} 
g - b g^{-1} b & b g^{-1} \\ 
- g^{-1} b & g^{-1} \end{pmatrix}
\in O(D,D)~, 
\label{GeneralizedMetric}
\end{equation}
define the $O(D,D)$ invariant metric and the generalized metric, with 
$\mathcal{H}\in O(D,D)$, resp.
Hence, Tseytlin's approach \cite{t90a,t90b} makes T--duality covariance manifest.

Tseytlin's formulation can straightforwardly be generalized to cases where the target space does not posses any isometries anymore: In addition to arbitrary non--constant symmetric matrices $\mathcal{G}$ (replacing $\eta$) and $\mathcal{H}$, there can be an anti--symmetric tensor $\mathcal{C}$. In the process of this generalization Lorentz invariance is lost. This can be seen by representing the classical Tseytlin's form as 
\begin{equation}
S_\text{cl.} =  \int\! 
 \Big\{ 
\frac 12\, \partial_L Y^T \big( \mathcal {H} + \mathcal{C} \big) \partial_R Y 
- \frac 12\, W_+^T \mathcal{G} W_-
\Big\}~, 
\label{ProjAction}
\end{equation}
where $W_\pm = \Pi_+ \partial_L Y \pm \Pi_- \partial_R Y$ with
$\Pi_\pm = \frac 12 \big (1 \pm  \mathcal{G}^{-1} \mathcal{H}\big)$. 
In order to ensure that $D$ of $2D$ d.o.f.'s are removed and the Lorentz invariance is restored, one may demand \cite{t90b} that  $W_+=0$ and that the operators $\Pi_\pm$ are projectors (which requires that $\mathcal{H} \in O(D,D)$ w.r.t.\ $\mathcal{G}$, i.e.\ $\mathcal{H}^T \mathcal{G}^{-1} \mathcal{H} = \mathcal{G}$), hence: $\Pi_+ \partial_L Y = \Pi_- \partial_R Y = 0$. 
In the Tseytlin's formulation one has to impose that both Lorentz and Weyl invariance are preserved at the quantum level \cite{Sfetsos:2009vt,Avramis:2009xi}. This leads to two conditions, which obscure the identification of the worldsheet beta functions with the target space e.o.m.'s.

\section{Lorentz invariant Tseytlin formalism}

\subsection{Construction} 

We would like to obtain a Tseytlin--like formulation which admits that the action is parameterized by arbitrary matrix valued functions and which is manifestly Lorentz invariant at the same time. The reason that the Tseytlin approach is not (manifestly) Lorentz invariant is easily identified in the construction above: A Lorentz breaking axial gauge fixing was used. Hence the remedy is to impose a Lorentz covariant gauge fixing.

A Lorentz preserving gauge fixing reads: $G=V_L$.  
Contrary to the axial gauge in Tseytlin's approach, after integrating over $\tilde B$ and $V_L$ the remaining gauge field component $V_R$ appears only linearly in the action
\begin{equation}\label{ActionVLnull}
S = \int\! 
\Big\{ 
\partial_L X^T E\, \partial_R X  + W_L^T V_R
- \partial_L \textsc{c}^T \textsc{b} 
\Big\}~, 
\end{equation}
hence it classically enforces the constraint 
\begin{equation}
W_L^T = \partial_L X^T  E + \partial_L \tilde X^T  
\stackrel{!}{=} 0~. 
\label{constraint0} 
\end{equation}
After a change of variables, $\tilde X \rightarrow \tilde X - E^T\, X$, the path integral contains a factor,  
\begin{equation}
Z_\text{ch.bos.} = 
\int\!\mathcal{D}[\tilde X, V_R]\, 
\exp i\!\int\! V_R^T \partial_L \tilde X~, 
\end{equation}
which might be viewed as a problematic chiral boson, see e.g.\ a comment \cite{Harada:1990fn} on \cite{Srivastava:1989zp}. However,  in the path integral corresponding to the action \eqref{ActionVLnull} this chiral boson contribution is precisely cancelled by the ghost sector.

\subsection{General form}

To go beyond constant backgrounds we take \eqref{ActionVLnull} at face value: It has the form of a gauge fixed action where the constraint \eqref{constraint0} defines the gauge condition. Then given the form of the ghost action in \eqref{ActionVLnull}, the gauge transformation being fixed can be inferred to be e.g.: 
\begin{equation} 
\delta \tilde{X} = \tilde{\xi}~. 
\label{deltatX} 
\end{equation} 
This gauge transformation could have been anticipated from Buscher's gauge theory \eqref{PathInt}: The pure gauge requirement is equally well implemented when $\tilde{X} \rightarrow \tilde{X}+\tilde{\xi}$.

To generalize to non--constant configurations we simply ask for a generic kinetic action compatible with this gauge transformation and a generic gauge fixing term. 
Since \eqref{deltatX} transforms $\tilde{X}$ with a shift, the most general kinetic term is as in \eqref{Polyakov} while the constraint 
\begin{equation}
W_L^T = \partial_L X^T  K(Y) + \partial_L \tilde X^T L(Y) 
\label{constraint} 
\end{equation}
may involve general $D\times D$ matrix functions $K, L$ of $Y$. Therefore, we take the classical action 
\begin{equation}
S_\text{cl.} =  \int\! 
\Big\{ 
\partial_L X^T E(X)\partial_R X + 
W_L^T  V_R 
\Big\}~, 
\label{GeneralActionII}
\end{equation}
as our starting point. If we retain the interpretation of \eqref{constraint} as  a gauge fixing condition for coordinate redundancy,  its variation determines the ghost action to be 
\begin{equation*}
S_\text{gh.} = - \int\! \Big\{ \partial_L \textsc{c}^T L
+ \partial_L X^T K_,{}^\mu \textsc{c}_\mu 
+ \partial_L \tilde{X}^T L_,{}^\mu \textsc{c}_\mu 
\Big\} 
\textsc{b}~. 
\end{equation*}
The matrices $K$ and $L$  are only defined up to an invertible matrix function $\rho$: 
\begin{equation}
K \rightarrow K\, \rho^{-1}~, 
\quad 
L \rightarrow L\, \rho^{-1}~, 
\quad 
V_R \rightarrow \rho \, V_R~. 
\label{RedefineVR}
\end{equation}
Since the matrices $K, L$ only appear in the gauge fixing term, we do not interpret them as physical target space d.o.f.'s. Hence, this theory describes $D^2$ d.of.'s as the standard sigma--model.

Using the constraint \eqref{constraint} the classical action can be represented in various guises: By the transformation 
\begin{equation}
V_R \rightarrow V_R 
+ \kappa\, \partial_R X + \lambda\, \partial_R \tilde{X}~, 
\label{ShiftVR}
\end{equation}
with matrix functions $\kappa$ and $\lambda$, 
the action \eqref{GeneralActionII} becomes 
\begin{equation}
S_\text{cl.} =  \int  
\Big\{
\frac 12\, \partial_L Y^T 
\big( \mathcal{G} + \mathcal{C} \big) 
\partial_R Y + 
\partial_L Y^T 
\begin{pmatrix} K \\ L \end{pmatrix}
V_R
\Big\}
\label{GeneralAction} 
\end{equation}
where 
\begin{equation}
\frac12 \big(\mathcal{G} + \mathcal{C}\big) 
=  
\begin{pmatrix}
E + K\, \kappa & K \,\lambda -\mu  \\
L\, \kappa+\mu & L\, \lambda 
\end{pmatrix}~. 
\label{UsingConstraint}
\end{equation}
Any constant matrix $\mu$ can be introduced here by a double partial integration 
$\partial_L \tilde X^T\mu\, \partial_R X = \partial_L X^T\mu\, \partial_R \tilde X$.

\subsection{Standard sigma--model \textbf{\textit{O(D,D)}} covariance}

Using \eqref{constraint} the classical action can be written as
\begin{equation}\label{actionA}
S_\text{cl.} = \int 
\Big\{ 
 - \frac 12\, \partial_L Y^T \eta\, \partial_R Y 
 + \partial_L Y^T  \begin{pmatrix} K \\ L \end{pmatrix} V_R
 \Big\}
\end{equation}
with $K=E$ and $L=\mathbbm{1}$. We brought the quadratic part of the action into a form involving $\eta$, given in \eqref{GeneralizedMetric}, by employing a double partial integration as mentioned above 
(i.e.\ we took $\kappa = -\mathbbm{1}$, $\lambda=0$ and
$\mu = \frac 12 \mathbbm{1}$ in \eqref{UsingConstraint}). 
The action \eqref{actionA} is 
invariant under global $O(D,D)$ mappings,  
%
\begin{equation}
Y \rightarrow Y' = M^{-T} Y~, 
\quad 
M = \begin{pmatrix} \alpha & \beta \\ \gamma & \delta \end{pmatrix} 
\in O(D,D)~,
\end{equation}
provided that they are extended to 
\begin{equation}
\begin{pmatrix} K \\ L \end{pmatrix} 
\rightarrow  \begin{pmatrix} K' \\ L' \end{pmatrix} 
=  M \begin{pmatrix} K \\ L \end{pmatrix}~,
\quad 
V_R' = V_R~.
\label{LinearTransEI}
\end{equation}
In fact, for $K=E$, $L=L'=\mathbbm{1}$ can be kept inert provided that the Lagrange multiplier $V_R$ transforms (see \eqref{RedefineVR}) as 
\begin{equation}\label{transVR}
V_R \rightarrow \big(\gamma E(Y) + \delta \big)\, V_R~, 
\end{equation}
which induces a fractional linear transformation 
\begin{equation}
E(Y) \rightarrow 
E'(Y') = 
\big(\alpha E(Y) + \beta\big)
\big(\gamma E(Y) + \delta\big)^{-1}~. 
\label{FractionalLinear}
\end{equation}

The action \eqref{actionA} can equivalently be represented in a form closely related to Hull's doubled formalism \cite{Hull:2004in,Hull:2006va}: 
\begin{equation}
S_\text{cl.} = \int  
\Big\{  
\frac{1}{2} \partial_L Y^T \mathcal{H}(Y)\, \partial_R Y
 + W_L(Y)^T  V_R
\Big\}~, 
 \label{actionH} 
\end{equation}
in terms of the generalized metric $\mathcal{H}(Y)$, see \eqref{GeneralizedMetric}. It follows that  
\begin{equation}
\mathcal{H}(Y) \rightarrow 
\mathcal{H}'(Y') = M \,\mathcal{H}(Y)\, M^T~. 
\end{equation}
All $O(D,D)$ transformations have been obtained in our formalism without referring to any isometries and coincide with the transformation rules in e.g.\ \cite{Hull:2009mi,Hohm:2010pp}.

\subsection{Geometric and non--geometric fluxes}

As the constraint is not invariant under $b$--field gauge transformations, fluxes have to be encoded in the quadratic terms. The  field strength \cite{Dall'Agata:2008qz} 
\begin{equation}
\mathcal{F}_3 = d \mathcal{C}_2
\label{FieldStrengthC}
\end{equation}
is left invariant by $\delta \mathcal{C}_2(Y) = d \Xi_1(Y)$ and hence generalizes the $b$--field gauge transformation. 

This suggests that fluxes should be introduced by identifying $b$ as components of $\mathcal{C}$ and not as part of the constraint. Motivated by the index structure, the components of $\mathcal{F}$, can be identified with the $H$--, geometric $f$-- and non--geometric $Q$-- and $R$--fluxes:
\begin{equation}
\begin{matrix} 
H_{\mu\nu\rho} = \mathcal{F}_{\mu\nu\rho}~,  
\quad 
f^\mu{}_{\nu\rho} = \mathcal{F}^\mu{}_{\nu\rho}~, 
\\[1ex] 
Q^{\mu\nu}{}_\rho = \mathcal{F}^{\mu\nu}{}_\rho~,
\quad 
R^{\mu\nu\rho} = \mathcal{F}^{\mu\nu\rho}~. 
\end{matrix} 
\label{Fluxes} 
\end{equation}

\section{Example: torus with linear \textbf{\textit{b}}--field}

Consider a three dimensional torus $T^3$ with periodicities 
$T_x:~x \rightarrow x+1$, etc, with a flat metric $g= \mathbbm{1}$ and a linear $b$--field in the $x,y$ directions, e.g.\ 
\begin{equation}\label{defB}
b(z) = z\, \omega~, 
\quad 
\omega = 
\frac 12 \big( e_x e_y^T - e_y e_x^T \big)~, 
\end{equation}
The $e_i$'s are the standard unit vectors with  
$e_i^T e_j = \delta_{ij}$. There are two possibilities to implement this background in the framework presented above:

\subsection{\textbf{\textit{b}}--field encoded in the constraint}

A description, which has manifest $O(3,3)$ invariant kinetic terms, is provided by \eqref{actionA} with $K= 1 +b(z)$, $L=\mathbbm{1}$. This action is invariant under the torus periodicities $T_x$ and $T_y$. However, since $T_z: b(z) \rightarrow b(z) + \omega$, the periodicity in the $z$--direction can only be realized, if it is accompanied by an appropriate $O(3,3)$ transformation:
\begin{equation}
T_z:~ Y \rightarrow 
\begin{pmatrix} 
\mathbbm{1}  & 0  \\ 
\omega & \mathbbm{1} 
\end{pmatrix} 
Y + 
\begin{pmatrix} e_z \\ 0 \end{pmatrix}~. 
\end{equation}
Since the three--torus is doubled, we have in addition 
\begin{equation}
T_{j}:~ Y \rightarrow Y + 
\begin{pmatrix} e_j \\ 0 \end{pmatrix}~, 
\quad 
T_{\tilde \jmath}:~ Y \rightarrow Y + 
\begin{pmatrix} 0 \\ e_{\tilde \jmath} \end{pmatrix}~, 
\end{equation}
with $j = x,y$ and $\tilde \jmath = \tilde x, \tilde y, \tilde z$. (Similar boundary conditions were found in \cite{Hull:2009sg}.) Consequently not all torus periodicities commute: 
$[T_z, T_y] = \frac 12\, T_{\tilde x}$, 
$[T_z, T_x] = -\frac 12\, T_{\tilde y}$.

Standard T--duality transformations are part of the $O(D,D)$ transformations \eqref{FractionalLinear} by defining a T--duality in the $e_i$--direction as
\begin{equation}
M_i = \begin{pmatrix}
\mathbbm{1} - e_i e_i^T & e_i e_i^T 
\\[1ex]  
e_i e_i^T & \mathbbm{1} - e_i e_i^T 
\end{pmatrix}~. 
\end{equation}
Given that $M_i^2 = \mathbbm{1}$, we denote combined $T$--dualities in the $e_i$ and $e_j$ directions, $i \neq j$, by $M_{ij}=M_{ji}$; $M_{xyz}$ denotes T--duality in all three directions. By \eqref{LinearTransEI}, subsequent T--dualities on the constraint read
\begin{gather*}
\begin{pmatrix} K \\ L \end{pmatrix}  = 
\begin{pmatrix} \mathbbm{1} + b(z) \\ \mathbbm{1} \end{pmatrix}~,
\quad 
M_x 
\begin{pmatrix} K \\ L \end{pmatrix} 
= 
\begin{pmatrix} \mathbbm{1} - \frac 12 \,z\, e_y e_x^T  
\\[1ex]   
\mathbbm{1} + \frac 12\, z\, e_x e_y^T \end{pmatrix}~,
\\[1ex] 
M_{xy} \begin{pmatrix} K \\ L \end{pmatrix}  
= 
\begin{pmatrix} \mathbbm{1} \\  \mathbbm{1}+b(z) \end{pmatrix}~,
\quad 
M_{xyz} \begin{pmatrix} K \\ L  \end{pmatrix} 
= \begin{pmatrix} \mathbbm{1} \\  \mathbbm{1}+b(\tilde z) \end{pmatrix}~.
\nonumber
\end{gather*}
Normally, applying a Buscher T--duality in the third direction is considered to be meaningless because there is no isometry in the $z$--direction. In our formalism we do not encounter any such problem: By construction all transformations are always linear; only if we employ appropriate accompanying $V_R$ transformations \eqref{transVR}, we obtain non--linear expressions, i.e. non-trivial monodromies of the target space fields. Hence, the trouble with naively applying the T--duality rules is that the torus periodicities are then only realized up to $V_R$ transformations.

\subsection{Fluxes as components of the doubled field strength}

As explained, in order that the full $b$--field gauge transformations leave the theory inert and fluxes can be systematically defined, we use the description where the $b$--field is part of the tensor $\mathcal{C}_{mn}=b_{mn}(z)$. This means that the constraint, encoded by $K = L = \mathbbm{1}$, is $T$--duality invariant and consequently the periodicity generators, $T_j, T_{\tilde \jmath}$, all commute. Moreover, taking $\kappa = -\mathbbm{1}$, 
$\lambda=0$ and $\mu = \frac 12\mathbbm{1}$ the action \eqref{GeneralAction} has 
\begin{equation} 
\frac 12 \big(\mathcal{G}+ \mathcal{C}\big)  = - \frac 12\, \eta + 
\begin{pmatrix}
b & 0 \\ 0 & 0 
\end{pmatrix}~.  
\end{equation}
Using this description, the chain of T--dual fluxes \cite{stw05}, 
\begin{equation}
H_{123} \stackrel{M_x~}{\longrightarrow} 
f^1{}_{23} \stackrel{M_y~}{\longrightarrow} 
Q^{12}{}_3 \stackrel{M_z~}{\longrightarrow} R^{123}~, 
\end{equation}
can be recovered from \eqref{Fluxes}. Since $\mathcal{C}$ transforms covariantly under $O(3,3)$ transformations, the forms of $\mathcal{C}$ and its field strength $\mathcal{F}$ under subsequent T--dualities in the $x$, $y$ and $z$--directions are easily computed, which shows that only one of them is switched on (and constant) in each of the duality frames. 
\\

\section{Outlook}

In the light of moduli stabilization, it would be interesting to study other  examples where several types of fluxes are switched on simultaneously, e.g.\ certain WZW models \cite{Dall'Agata:2008qz}, Poisson geometries \cite{Halmagyi:2008dr} or asymmetric orbifolds \cite{Condeescu:2012sp}. For this it is  necessary to investigate the e.o.m.'s via the $\beta$-functions \cite{Sfetsos:2009vt,Avramis:2009xi,Copland:2011wx}.

\subsection{Acknowledgements}
We would like to thank 
David Andriot, 
Gianguido Dall'Agata, 
Andreas Deser and 
Olaf Hohm 
for very helpful discussions. 
We are indebted to the referees for questioning the appearance of chiral bosons. 
This work has been supported by the LMUExcellent Programme.

%
%
%


\begin{thebibliography}{21}%
\makeatletter
\providecommand \@ifxundefined [1]{%
 \@ifx{#1\undefined}
}%
\providecommand \@ifnum [1]{%
 \ifnum #1\expandafter \@firstoftwo
 \else \expandafter \@secondoftwo
 \fi
}%
\providecommand \@ifx [1]{%
 \ifx #1\expandafter \@firstoftwo
 \else \expandafter \@secondoftwo
 \fi
}%
\providecommand \natexlab [1]{#1}%
\providecommand \enquote  [1]{``#1''}%
\providecommand \bibnamefont  [1]{#1}%
\providecommand \bibfnamefont [1]{#1}%
\providecommand \citenamefont [1]{#1}%
\providecommand \href@noop [0]{\@secondoftwo}%
\providecommand \href [0]{\begingroup \@sanitize@url \@href}%
\providecommand \@href[1]{\@@startlink{#1}\@@href}%
\providecommand \@@href[1]{\endgroup#1\@@endlink}%
\providecommand \@sanitize@url [0]{\catcode `\\12\catcode `\$12\catcode
  `\&12\catcode `\#12\catcode `\^12\catcode `\_12\catcode `\%12\relax}%
\providecommand \@@startlink[1]{}%
\providecommand \@@endlink[0]{}%
\providecommand \url  [0]{\begingroup\@sanitize@url \@url }%
\providecommand \@url [1]{\endgroup\@href {#1}{\urlprefix }}%
\providecommand \urlprefix  [0]{URL }%
\providecommand \Eprint [0]{\href }%
\providecommand \doibase [0]{http://dx.doi.org/}%
\providecommand \selectlanguage [0]{\@gobble}%
\providecommand \bibinfo  [0]{\@secondoftwo}%
\providecommand \bibfield  [0]{\@secondoftwo}%
\providecommand \translation [1]{[#1]}%
\providecommand \BibitemOpen [0]{}%
\providecommand \bibitemStop [0]{}%
\providecommand \bibitemNoStop [0]{.\EOS\space}%
\providecommand \EOS [0]{\spacefactor3000\relax}%
\providecommand \BibitemShut  [1]{\csname bibitem#1\endcsname}%
\let\auto@bib@innerbib\@empty
\bibitem [{\citenamefont {Buscher}(1987)}]{b87a}%
  \BibitemOpen
  \bibfield  {author} {\bibinfo {author} {\bibfnamefont {T.}~\bibnamefont
  {Buscher}},\ }\href {\doibase 10.1016/0370-2693(87)90769-6} {\bibfield
  {journal} {\bibinfo  {journal} {Phys.Lett.}\ }\textbf {\bibinfo {volume}
  {B194}},\ \bibinfo {pages} {59} (\bibinfo {year} {1987})}\BibitemShut
  {NoStop}%
\bibitem [{\citenamefont {Buscher}(1988)}]{b87b}%
  \BibitemOpen
  \bibfield  {author} {\bibinfo {author} {\bibfnamefont {T.}~\bibnamefont
  {Buscher}},\ }\href {\doibase 10.1016/0370-2693(88)90602-8} {\bibfield
  {journal} {\bibinfo  {journal} {Phys.Lett.}\ }\textbf {\bibinfo {volume}
  {B201}},\ \bibinfo {pages} {466} (\bibinfo {year} {1988})}\BibitemShut
  {NoStop}%
\bibitem [{\citenamefont {Tseytlin}(1990)}]{t90a}%
  \BibitemOpen
  \bibfield  {author} {\bibinfo {author} {\bibfnamefont {A.A.}\ \bibnamefont
  {Tseytlin}},\ }\href {\doibase 10.1016/0370-2693(90)91454-J} {\bibfield
  {journal} {\bibinfo  {journal} {Phys.Lett.}\ }\textbf {\bibinfo {volume}
  {B242}},\ \bibinfo {pages} {163} (\bibinfo {year} {1990})}\BibitemShut
  {NoStop}%
\bibitem [{\citenamefont {Tseytlin}(1991)}]{t90b}%
  \BibitemOpen
  \bibfield  {author} {\bibinfo {author} {\bibfnamefont {A.A.}\ \bibnamefont
  {Tseytlin}},\ }\href {\doibase 10.1016/0550-3213(91)90266-Z} {\bibfield
  {journal} {\bibinfo  {journal} {Nucl.Phys.}\ }\textbf {\bibinfo {volume}
  {B350}},\ \bibinfo {pages} {395} (\bibinfo {year} {1991})}\BibitemShut
  {NoStop}%
\bibitem [{\citenamefont {Floreanini}\ and\ \citenamefont
  {Jackiw}(1987)}]{Floreanini:1987as}%
  \BibitemOpen
  \bibfield  {author} {\bibinfo {author} {\bibfnamefont {R.}~\bibnamefont
  {Floreanini}}\ and\ \bibinfo {author} {\bibfnamefont {R.}~\bibnamefont
  {Jackiw}},\ }\href {\doibase 10.1103/PhysRevLett.59.1873} {\bibfield
  {journal} {\bibinfo  {journal} {Phys.Rev.Lett.}\ }\textbf {\bibinfo {volume}
  {59}},\ \bibinfo {pages} {1873} (\bibinfo {year} {1987})}\BibitemShut
  {NoStop}%
%
\bibitem [{\citenamefont {Duff}\ and\ \citenamefont
  {Duff}(1989)}]{Duff:1989tf}%
  \BibitemOpen
  \bibfield  {author} {\bibinfo {author} {\bibfnamefont {M.J.}~\bibnamefont
  {Duff}},\ }\href {\doibase 10.1016/0550-3213(90)90520-N} {\bibfield
  {journal} {\bibinfo  {journal} {Nucl.Phys.}\ }\textbf {\bibinfo {volume}
  {B335}},\ \bibinfo {pages} {610} (\bibinfo {year} {1989})}\BibitemShut
  {NoStop}%
%
\bibitem [{\citenamefont {Dall'Agata}\ and\ \citenamefont
  {Prezas}(2008)}]{Dall'Agata:2008qz}%
  \BibitemOpen
  \bibfield  {author} {\bibinfo {author} {\bibfnamefont {G.}~\bibnamefont
  {Dall'Agata}}\ and\ \bibinfo {author} {\bibfnamefont {N.}~\bibnamefont
  {Prezas}},\ }\href {\doibase 10.1088/1126-6708/2008/08/088} {\bibfield
  {journal} {\bibinfo  {journal} {JHEP}\ }\textbf {\bibinfo {volume} {0808}},\
  \bibinfo {pages} {088} (\bibinfo {year} {2008})} \BibitemShut
  {NoStop}%
%
\bibitem [{\citenamefont {Sfetsos}\ \emph {et~al.}(2010)\citenamefont
  {Sfetsos}, \citenamefont {Siampos},\ and\ \citenamefont
  {Thompson}}]{Sfetsos:2009vt}%
  \BibitemOpen
  \bibfield  {author} {\bibinfo {author} {\bibfnamefont {K.}\ \bibnamefont
  {Sfetsos}}, \bibinfo {author} {\bibfnamefont {K.}\ \bibnamefont
  {Siampos}}, \ and\ \bibinfo {author} {\bibfnamefont {D.C.}~\bibnamefont
  {Thompson}},\ }\href {\doibase 10.1016/j.nuclphysb.2009.11.001} {\bibfield
  {journal} {\bibinfo  {journal} {Nucl.Phys.}\ }\textbf {\bibinfo {volume}
  {B827}},\ \bibinfo {pages} {545} (\bibinfo {year} {2010})} \BibitemShut
  {NoStop}%
%
\bibitem [{\citenamefont {Avramis}\ \emph {et~al.}(2010)\citenamefont
  {Avramis}, \citenamefont {Derendinger},\ and\ \citenamefont
  {Prezas}}]{Avramis:2009xi}%
  \BibitemOpen
  \bibfield  {author} {\bibinfo {author} {\bibfnamefont {S.D.}\ \bibnamefont
  {Avramis}}, \bibinfo {author} {\bibfnamefont {J.-P.}\ \bibnamefont
  {Derendinger}}, \ and\ \bibinfo {author} {\bibfnamefont {N.}~\bibnamefont
  {Prezas}},\ }\href {\doibase 10.1016/j.nuclphysb.2009.11.003} {\bibfield
  {journal} {\bibinfo  {journal} {Nucl.Phys.}\ }\textbf {\bibinfo {volume}
  {B827}},\ \bibinfo {pages} {281} (\bibinfo {year} {2010})} \BibitemShut
  {NoStop}%
\bibitem [{\citenamefont {Hull}(2005)}]{Hull:2004in}%
  \BibitemOpen
  \bibfield  {author} {\bibinfo {author} {\bibfnamefont {C.}~\bibnamefont
  {Hull}},\ }\href {\doibase 10.1088/1126-6708/2005/10/065} {\bibfield
  {journal} {\bibinfo  {journal} {JHEP}\ }\textbf {\bibinfo {volume} {0510}},\
  \bibinfo {pages} {065} (\bibinfo {year} {2005})}
  \BibitemShut {NoStop}%
\bibitem [{\citenamefont {Hull}(2007)}]{Hull:2006va}%
  \BibitemOpen
  \bibfield  {author} {\bibinfo {author} {\bibfnamefont {C.}\ \bibnamefont
  {Hull}},\ }\href {\doibase 10.1088/1126-6708/2007/07/080} {\bibfield
  {journal} {\bibinfo  {journal} {JHEP}\ }\textbf {\bibinfo {volume} {0707}},\
  \bibinfo {pages} {080} (\bibinfo {year} {2007})}
  \BibitemShut {NoStop}%
\bibitem [{\citenamefont {Berman}\ and\ \citenamefont
  {Thompson}(2008)}]{Berman:2007yf}%
  \BibitemOpen
  \bibfield  {author} {\bibinfo {author} {\bibfnamefont {D.S.}\ \bibnamefont
  {Berman}}\ and\ \bibinfo {author} {\bibfnamefont {D.C.}\ \bibnamefont
  {Thompson}},\ }\href {\doibase 10.1016/j.physletb.2008.03.012} {\bibfield
  {journal} {\bibinfo  {journal} {Phys.Lett.}\ }\textbf {\bibinfo {volume}
  {B662}},\ \bibinfo {pages} {279} (\bibinfo {year} {2008})} \BibitemShut
  {NoStop}%
\bibitem [{\citenamefont {Berman}\ \emph {et~al.}(2008)\citenamefont {Berman},
  \citenamefont {Copland},\ and\ \citenamefont {Thompson}}]{Berman:2007xn}%
  \BibitemOpen
  \bibfield  {author} {\bibinfo {author} {\bibfnamefont {D.S.}\ \bibnamefont
  {Berman}}, \bibinfo {author} {\bibfnamefont {N.B.}\ \bibnamefont {Copland}},
  \ and\ \bibinfo {author} {\bibfnamefont {D.C.}\ \bibnamefont {Thompson}},\
  }\href {\doibase 10.1016/j.nuclphysb.2007.09.021} {\bibfield  {journal}
  {\bibinfo  {journal} {Nucl.Phys.}\ }\textbf {\bibinfo {volume} {B791}},\
  \bibinfo {pages} {175} (\bibinfo {year} {2008})} \BibitemShut
  {NoStop}%
\bibitem [{\citenamefont {Pasti}\ \emph {et~al.}(1997)\citenamefont {Pasti},
  \citenamefont {Sorokin},\ and\ \citenamefont {Tonin}}]{Pasti:1996vs}%
  \BibitemOpen
  \bibfield  {author} {\bibinfo {author} {\bibfnamefont {P.}~\bibnamefont
  {Pasti}}, \bibinfo {author} {\bibfnamefont {D.P.}\ \bibnamefont {Sorokin}},
  \ and\ \bibinfo {author} {\bibfnamefont {M.}~\bibnamefont {Tonin}},\ }
  \href{\doibase 10.1103/PhysRevD.55.6292} 
  {\bibfield  {journal} {\bibinfo
  {journal} {Phys.Rev.}\ }\textbf {\bibinfo {volume} {D55}},\ \bibinfo {pages}
  {6292} (\bibinfo {year} {1997})}
  \BibitemShut {NoStop}%
\bibitem [{\citenamefont {Callan}\ \emph {et~al.}(1985)\citenamefont {Callan},
  \citenamefont {Martinec}, \citenamefont {Perry},\ and\ \citenamefont
  {Friedan}}]{Callan:1985ia}%
  \BibitemOpen
  \bibfield  {author} {\bibinfo {author} {\bibfnamefont {J.}~\bibnamefont
  {Callan}, \bibfnamefont {et~al.}}
  ,\ }\href {\doibase 10.1016/0550-3213(85)90506-1}
  {\bibfield  {journal} {\bibinfo  {journal} {Nucl.Phys.}\ }\textbf {\bibinfo
  {volume} {B262}},\ \bibinfo {pages} {593} (\bibinfo {year}
  {1985})}\BibitemShut {NoStop}%
\bibitem [{\citenamefont {Hull}\ and\ \citenamefont
  {Townsend}(1986)}]{Hull:1985rc}%
  \BibitemOpen
  \bibfield  {author} {\bibinfo {author} {\bibfnamefont {C.}~\bibnamefont
  {Hull}}\ and\ \bibinfo {author} {\bibfnamefont {P.}~\bibnamefont
  {Townsend}},\ }\href {\doibase 10.1016/0550-3213(86)90289-0} {\bibfield
  {journal} {\bibinfo  {journal} {Nucl.Phys.}\ }\textbf {\bibinfo {volume}
  {B274}},\ \bibinfo {pages} {349} (\bibinfo {year} {1986})}\BibitemShut
  {NoStop}%
\bibitem [{\citenamefont {Hull}\ and\ \citenamefont
  {Zwiebach}(2009)}]{Hull:2009mi}%
  \BibitemOpen
  \bibfield  {author} {\bibinfo {author} {\bibfnamefont {C.}~\bibnamefont
  {Hull}}\ and\ \bibinfo {author} {\bibfnamefont {B.}~\bibnamefont
  {Zwiebach}},\ }\href {\doibase 10.1088/1126-6708/2009/09/099} {\bibfield
  {journal} {\bibinfo  {journal} {JHEP}\ }\textbf {\bibinfo {volume} {0909}},\
  \bibinfo {pages} {099} (\bibinfo {year} {2009})} \BibitemShut
  {NoStop}%
\bibitem [{\citenamefont {Rocek}\ and\ \citenamefont
  {Tseytlin}(1999)}]{Rocek:1997hi}%
  \BibitemOpen
  \bibfield  {author} {\bibinfo {author} {\bibfnamefont {M.}~\bibnamefont
  {Rocek}}\ and\ \bibinfo {author} {\bibfnamefont {A.A.}\ \bibnamefont
  {Tseytlin}},\ }\href {\doibase 10.1103/PhysRevD.59.106001} {\bibfield
  {journal} {\bibinfo  {journal} {Phys.Rev.}\ }\textbf {\bibinfo {volume}
  {D59}},\ \bibinfo {pages} {106001} (\bibinfo {year} {1999})}
  \BibitemShut {NoStop}%
\bibitem[{\citenamefont{Harada}(1990)}]{Harada:1990fn}
\bibinfo{author}{\bibfnamefont{K.}~\bibnamefont{Harada},\ }
  \href {\doibase 10.1103/PhysRevLett.65.267} 
  {\bibfield  {journal}
  \bibinfo{journal}{Phys.Rev.Lett.} \textbf{\bibinfo{volume}{65}},
  \bibinfo{pages}{267} (\bibinfo{year}{1990})}
   \BibitemShut {NoStop}%
\bibitem[{\citenamefont{Srivastava}(1989)}]{Srivastava:1989zp}
\bibinfo{author}{\bibfnamefont{P.~P.} \bibnamefont{Srivastava},\ }
 \href {\doibase 10.1103/PhysRevLett.63.2791} 
  {\bibfield  {journal}
  \bibinfo{journal}{Phys.Rev.Lett.} \textbf{\bibinfo{volume}{63}},
  \bibinfo{pages}{2791} (\bibinfo{year}{1989})}
 \BibitemShut {NoStop}%
\bibitem [{\citenamefont {Hohm}\ \emph {et~al.}(2010)\citenamefont {Hohm},
  \citenamefont {Hull},\ and\ \citenamefont {Zwiebach}}]{Hohm:2010pp}%
  \BibitemOpen
  \bibfield  {author} {\bibinfo {author} {\bibfnamefont {O.}~\bibnamefont
  {Hohm}}, \bibinfo {author} {\bibfnamefont {C.}~\bibnamefont {Hull}}, \ and\
  \bibinfo {author} {\bibfnamefont {B.}~\bibnamefont {Zwiebach}},\ }
  \href{\doibase 10.1007/JHEP08(2010)008} 
  {\bibfield  {journal} {\bibinfo  {journal}
  {JHEP}\ }\textbf {\bibinfo {volume} {1008}},\ \bibinfo {pages} {008}
  (\bibinfo {year} {2010})} \BibitemShut {NoStop}%
\bibitem [{\citenamefont {Hull}\ and\ \citenamefont
  {Reid-Edwards}(2009)}]{Hull:2009sg}%
  \BibitemOpen
  \bibfield  {author} {\bibinfo {author} {\bibfnamefont {C.}~\bibnamefont
  {Hull}}\ and\ \bibinfo {author} {\bibfnamefont {R.}~\bibnamefont
  {Reid-Edwards}},\ }\href {\doibase 10.1088/1126-6708/2009/09/014} {\bibfield
  {journal} {\bibinfo  {journal} {JHEP}\ }\textbf {\bibinfo {volume} {0909}},\
  \bibinfo {pages} {014} (\bibinfo {year} {2009})} \BibitemShut
  {NoStop}%
\bibitem [{\citenamefont {Shelton}\ \emph {et~al.}(2005)\citenamefont
  {Shelton}, \citenamefont {Taylor},\ and\ \citenamefont {Wecht}}]{stw05}%
  \BibitemOpen
  \bibfield  {author} {\bibinfo {author} {\bibfnamefont {J.}~\bibnamefont
  {Shelton}}, \bibinfo {author} {\bibfnamefont {W.}~\bibnamefont {Taylor}}, \
  and\ \bibinfo {author} {\bibfnamefont {B.}~\bibnamefont {Wecht}},\ }
  \href{\doibase 10.1088/1126-6708/2005/10/085} 
  {\bibfield  {journal} {\bibinfo
  {journal} {JHEP}\ }\textbf {\bibinfo {volume} {0510}},\ \bibinfo {pages}
  {085} (\bibinfo {year} {2005})}
  \BibitemShut {NoStop}%
\bibitem{Halmagyi:2008dr}%
  \BibitemOpen
  \bibfield{author}{%
  \bibinfo {author} {\bibfnamefont{N.}~\bibnamefont{Halmagyi}},\ }%
\href  {\doibase 10.1088/1126-6708/2008/07/137}
  {\bibfield{journal}{{\bibinfo {journal} {JHEP}}\ }%
  \textbf{\bibinfo {volume} {0807}},\ \bibinfo {pages} {137} (\bibinfo {year}
  {2008})}
  \BibitemShut {NoStop}%
\bibitem [{\citenamefont {Condeescu}\ \emph {et~al.}(2012)\citenamefont
  {Condeescu}, \citenamefont {Florakis},\ and\ \citenamefont
  {L\"ust}}]{Condeescu:2012sp}%
  \BibitemOpen
  \bibfield  {author} {\bibinfo {author} {\bibfnamefont {C.}~\bibnamefont
  {Condeescu}}, \bibinfo {author} {\bibfnamefont {I.}~\bibnamefont {Florakis}},
  \ and\ \bibinfo {author} {\bibfnamefont {D.}~\bibnamefont {L\"ust}},\ }
  \href {\doibase 10.1007/JHEP04(2012)121} 
  {\bibfield  {journal} 
  {\bibinfo  {journal}
  {JHEP}\ }\textbf {\bibinfo {volume} {1204}},\ \bibinfo {pages} {121}
  (\bibinfo {year} {2012})} \BibitemShut {NoStop}%
\bibitem [{\citenamefont {Copland}(2012)}]{Copland:2011wx}%
  \BibitemOpen
  \bibfield  {author} {\bibinfo {author} {\bibfnamefont {N.B.}\ \bibnamefont
  {Copland}},\ }
  \href {\doibase 10.1007/JHEP04(2012)044} 
  {\bibfield  {journal}
  {\bibinfo  {journal} {JHEP}\ }\textbf {\bibinfo {volume} {1204}},\ \bibinfo
  {pages} {044} (\bibinfo {year} {2012})} \BibitemShut
  {NoStop}%
\end{thebibliography}
\end{document}